\documentclass[pra,twocolumn,showpacs]{revtex4}
\usepackage{amsmath}
\usepackage{amsfonts}
\usepackage{amssymb}
\usepackage{graphicx}

\begin{document}

\title{Spin lattices with two-body Hamiltonians for which the ground state encodes a cluster state}
\author{Tom Griffin}%
\author{Stephen D. Bartlett}
\affiliation{School of Physics, The University of Sydney, Sydney,
New South Wales 2006, Australia}%

\date{4 December 2008}

\begin{abstract}
We present a general procedure for constructing lattices of qubits with a Hamiltonian composed of nearest-neighbour
two-body interactions such that the ground state encodes a cluster state.  We give specific details for lattices in one-, two-, and three-dimensions, investigating both periodic and fixed boundary conditions, as well as present a proof for the applicability of this procedure to any graph.  We determine the energy gap of these systems, which is shown to be independent of the size of the lattice but dependent on the type of lattice (in particular, the coordination number), and investigate the scaling of this gap in terms of the coupling constants of the Hamiltonian.  We provide a comparative analysis of the different lattice types with respect to their usefulness for measurement-based quantum computation.
\end{abstract}


\maketitle

\section{Introduction}

There is currently considerable interest in preparing exotic quantum states of many-body systems which can be used as resource states for measurement-based quantum computation (MBQC) -- that is, quantum computation that proceeds solely through local adaptive measurements on single quantum systems~\cite{Rau1,Rau3,Gro07a,Gro07b,vdN07,Bre08}.  The canonical example of such a resource state is the cluster state~\cite{Rau1,Rau3}, which is a universal resource for MBQC on suitable lattices or graphs~\cite{vdN06}.  It may be possible to prepare such a cluster state dynamically in atomic systems such as an optical lattice~\cite{Tre06} or using single photons~\cite{Bro05,Wal05}.  However, one exciting possibility is that such resource states might be the non-degenerate ground state of a ``natural'' Hamiltonian lattice system.  If the system is gapped, then one simply needs to cool it sufficiently in order to obtain the desired state (although, even for gapped systems, this cooling process may be difficult~\cite{Sch08}).

Consider the cluster state on a lattice $\mathcal{L}$, defined as the unique $+1$ eigenstate of a set of stabilizer operators $S_\mu = X_\mu \otimes_{\nu\sim\mu}Z_\nu$, where $X_\mu$ ($Z_\mu$) is the Pauli $X$ ($Z$) operator at site $\mu$ and where $\nu\sim\mu$ denotes that $\nu$ is connected to $\mu$ by a bond in the lattice $\mathcal{L}$.   The Hamiltonian
\begin{equation}
  \label{eq:ClusterHamiltonian}
  H = - \Delta \sum_{\mu \in \mathcal{L}} S_\mu \,,
\end{equation}
with $\Delta$ an energy constant, has the cluster state as its unique ground state~\cite{Rau3}.  In addition, this system is gapped (the gap is $2\Delta$), and such a system can be cooled efficiently~\cite{Ver08}.  However, for any non-trivial lattice or graph, this Hamiltonian involves many-body interactions, as opposed to the two-body interactions that occur frequently in nature.

An obvious question, then, is whether it is possible to realize any given highly-entangled quantum state as the ground state of a Hamiltonian with only two-body interactions.  Haselgrove \textit{et al.}~\cite{Has03} proved that this is not possible in general, and Nielsen~\cite{Nie05} used this result to prove that a cluster state on a computationally universal (i.e., two-dimensional or higher) lattice cannot arise as the ground state of a Hamiltonian with only two-body interactions.  However, investigations into quantum complexity theory~\cite{Kem04,Oli05} have demonstrated that cluster states (and other such
states that are universal) can be \emph{approximated} by the ground state of a local two-body Hamiltonian.  The key idea is to make use of ``mediating'' ancilla qubits to create an effective many-body coupling out of two-body interactions.  The problem with such methods is that the detailed parameters in the perturbing Hamiltonian must be controlled with a precision that increases with the size of the system~\cite{vdN08}, making such approaches impractical for the task of creating cluster states on large lattices.

Using an alternate method based on the idea behind projected entangled pair states (PEPS)~\cite{Ver04}, Bartlett and Rudolph~\cite{Bar06} proved that it was possible to obtain a state that closely approximates an \emph{encoded} cluster state on a square lattice using a Hamiltonian with only two-body nearest-neighbour interactions.  In addition, they proved that MBQC can proceed using such an encoded resource state, still requiring only adaptive single-qubit measurements.

In this paper, we present a general method for constructing two-body nearest-neighbour Hamiltonian systems for which the ground state encodes a cluster state, based on the techniques of~\cite{Bar06}.  Our rigorous application of perturbation theory reveals errors in the calculation of the energy gap for the square lattice investigated in~\cite{Bar06} (although these errors do not affect their key result) and we provide a correct treatment of this case.  We also investigate the cluster state on a one-dimensional line (useful for illustration, as well as for its application as a quantum wire~\cite{Gro07b}), a hexagonal lattice in two-dimensions -- a universal MBQC resource with the best scaling of the energy gap in perturbation, and the cubic lattice in three-dimensions -- a resource state for which fault-tolerance thresholds have been found~\cite{Rau06,Rau07}.  We explicitly characterise the effects of fixed boundary conditions on the lattice, proving that such boundary conditions do not affect the main result.  Finally, we provide an outline of a proof that this method yields an encoded cluster state as the ground state on any graph.

\section{A PEPS Hamiltonian}

Our general method relies on the fact that the cluster state is simply represented as a projected entangled pair state (PEPS), also known as a valence-bond solid state.

\subsection{The PEPS representation of a cluster state}

The PEPS representation~\cite{Ver04} is a powerful and often compact method of describing the state of a many-body system.  Consider a regular lattice $\mathcal{L}$ of qubits, with coordination number $c$ (i.e., $c$ bonds connect each qubit to other sites on the lattice).  A PEP state on $\mathcal{L}$ can be constructed by assigning a pair of \emph{virtual} quantum systems of dimension $D$ to each bond on the lattice, each pair prepared in a maximally-entangled state, and then applying a projection $P$ to the $c$ virtual systems associated with each site.  The cluster state (and a wide variety of other states of interest) require only $D=2$ for their representation; in what follows, we restrict our attention to this case where the virtual systems are qubits.  In addition, we choose the maximally-entangled state of these virtual qubits to be the two-qubit cluster state
\begin{equation}
  |C_2\rangle=\tfrac{1}{\sqrt{2}}\bigl(|0\rangle|+\rangle+|1\rangle|-\rangle \bigr)\,,
\end{equation}
where $|\pm \rangle=\frac{1}{\sqrt{2}}(|0\rangle \pm |1\rangle)$.  With this convention, the cluster state has a simple PEPS
representation~\cite{Ver04} corresponding to the projection operator
\begin{equation}
  P_L=|0_L\rangle \langle00\ldots 0|+|1_L\rangle \langle11\ldots 1|\,,
\end{equation}
at each site, with $c$ zeros (ones) in $\langle00\ldots 0|$ ($\langle11\ldots 1|$), and the states $|0_L\rangle$ and $|1_L\rangle$ forming a basis for the resulting qubit at each site.

As an example, consider the PEPS representation of the cluster state on a two-dimensional square lattice.  There are four bonds emanating from every site in a square lattice, and so each site possesses four virtual qubits.  Virtual qubits connected by a bond are placed in the state $|C_2\rangle$, and then a projection $P_L=\otimes_{\rm sites}(|0_L\rangle \langle0000|+|1_L\rangle \langle1111|)$ is applied.  The resultant state, $|\phi\rangle=P_L \otimes_{\rm bonds}|C_2\rangle$ is a cluster state on the square lattice.

%\begin{figure}[h]
%\begin{center}
%\includegraphics[width=80mm]{VBSstate}
%   \caption{The PEPS construction on a 2D square lattice}\label{fig:VBS}
%\end{center}
%\end{figure}

\subsection{A two-body PEPS Hamiltonian}

The essential idea of the method presented in this paper is to mimic the PEPS construction procedure using a physical two-body Hamiltonian, wherein the ``virtual'' qubits are real physical systems and the resulting PEPS state is \emph{encoded} into logical qubits.
Consider a regular lattice.  Let $\mathcal{L}$ denote the set of sites, each with coordination number $c$.  We assign $c$ qubits to each site, and label with a double index $(\mu,i)$, with $\mu \in \mathcal{L}$ and $i=1,2,\ldots,c$.  (The choice of the second label $i$ is completely arbitrary.) Let $\sigma^x_{(\mu,i)}$ and $\sigma^z_{(\mu,i)}$ denote the Pauli $X$ and $Z$ operators for the $i$th qubit at site $\mu \in \mathcal{L}$.  Following the PEPS construction, if a site $\mu$ is connected to a site $\nu$ by a bond (denoted $\mu \sim \nu$), then we associate qubit $(\mu,i)$ and $(\nu,i)$ for some $i$ to this bond.  (We note that this notation can become problematic with certain periodic conditions, but it should be clear from the context how to adjust it appropriately.)

Our \emph{PEPS Hamiltonian} is defined as follows.  At each site, we require a two-body Hamiltonian with a two-dimensional ground state space spanned by $|00\ldots 0\rangle$ and $|11\ldots 1\rangle$.  For this, we choose a site Hamiltonian $H_0$ which is of Ising form
\begin{equation}
  \label{eq:Ising}
  H_0 = - \sum_{\mu \in \mathcal{L}} \sum_{i\leftrightarrow j} \sigma^z_{(\mu,i)}\otimes \sigma^z_{(\mu,j)}\,,
\end{equation}
where $i \leftrightarrow j$ denotes that qubits $i$ and $j$ are connected according to some graph structure.  Aside from being connected, the specific form of this graph is relatively unimportant; however its structure will affect the energy levels of the Hamiltonian.  For example, for two-dimensional lattices, it is natural to choose a ring structure.

Between sites, we define a different two-body interaction of the form
\begin{equation}
  \label{eq:Bonds}
  V=-\sum_{\mu \in \mathcal{L}} \sum_{i=1}^c  \sigma^x_{(\mu,i)}\otimes\sigma^z_{(\nu(i),i)} \,,
\end{equation}
where $\nu(i)$ is the site connected to $\mu$ via bond $i$.  With this Hamiltonian, on every bond in the lattice $\mu \sim \nu$ there are two terms: $\sigma^x_{(\mu,i)} \otimes \sigma^z_{(\nu,i)}$ and $\sigma^z_{(\mu,i)}\otimes\sigma^x_{(\nu,i)}$.  Note that the terms in $V$ stabilise $|C_2\rangle$, and therefore $\otimes_{\rm bonds}|C_2\rangle$ is the ground state of $V$.  This product of maximally-entangled states is the starting point of the PEPS construction.  The site Hamiltonian $H_0$ is meant to ``implement'' the PEPS projection by ensuring that the qubits at a site act as a single logical qubit; to do so, the site Hamiltonian $H_0$ must be much stronger than the bond Hamiltonian $V$.  One is then lead to consider the ground state of the Hamiltonian \begin{equation}
  \label{hamilt}
  H=g H_0+\lambda V \,,
\end{equation}
where $g \gg \lambda>0$, which is suitable for perturbative analysis in $\lambda / g$.

In \cite{Bar06}, this procedure was applied to a square lattice. The terms in the perturbation combine at higher orders to yield the stabilisers of the logical cluster state, and the resulting low-energy theory of the lattice is governed by an effective Hamiltonian of the form of Eq.~(\ref{eq:ClusterHamiltonian}).  Furthermore, the gap to the next excited state is finite and independent of the size of the lattice. Of course, because this is a perturbative approach there will now be corrections to the unperturbed logical eigenstates which will not be in the logical space. So the \emph{exact} cluster state will not be obtained. However, these errors will be small (occurring with probability $(\lambda / g)^2$, as we will show) and so a state arbitrarily close to the cluster state can be obtained. In what follows, this procedure is generalised to other lattice structures important in quantum computation.

\subsection{General properties of the PEPS Hamiltonian}

\subsubsection{Duality transformation to uncoupled sites}
\label{subsec:ExactSol}

We now present a simple duality transformation that maps the Hamiltonian~\eqref{hamilt} to one describing uncoupled sites.  Consider decomposing~\eqref{hamilt} as a sum of commuting terms $H_\mu$, as
\begin{equation}
  H = \sum_{\mu \in \mathcal{L}} H_\mu\,,
\end{equation}
where
\begin{equation}
  H_\mu = - g \sum_{i\leftrightarrow j} \sigma^z_{(\mu,i)}\otimes \sigma^z_{(\mu,j)} - \lambda \sum_{i=1}^c  \sigma^x_{(\mu,i)}\otimes\sigma^z_{(\nu(i),i)} \,.
\end{equation}
Note that $[H_\mu,H_\nu]=0$ for $\mu\neq \nu$.  Define the unitary transformation $\text{CS}_{\mathcal{L}}$ to be the application of a $\text{CSIGN}$ gate
\begin{equation}
  \text{CSIGN}: \begin{cases} \sigma^x \otimes I &\rightarrow \sigma^x \otimes \sigma^z \\
    \sigma^z \otimes I &\rightarrow \sigma^z \otimes I \\
    I \otimes \sigma^x &\rightarrow \sigma^z \otimes \sigma^x \\
    I \otimes \sigma^z &\rightarrow I \otimes \sigma^z \end{cases}
\end{equation}
to every bond on the lattice.  The action of this transformation on the bond terms in the above Hamiltonian is $(\text{CS}_{\mathcal{L}})\sigma^x_{(\mu,i)}\otimes\sigma^z_{(\nu(i),i)}(\text{CS}_{\mathcal{L}}) = \sigma^x_{(\mu,i)}$.  Transforming each term $H_\mu$ thus yields
\begin{align}
  H_\mu' &= (\text{CS}_{\mathcal{L}})H_\mu(\text{CS}_{\mathcal{L}}) \nonumber \\
  &= - g \sum_{i\leftrightarrow j} \sigma^z_{(\mu,i)}\otimes \sigma^z_{(\mu,j)} - \lambda \sum_{i=1}^c  \sigma^x_{(\mu,i)} \,,
  \label{eq:TFIM}
\end{align}
which is localized entirely to the site $\mu$.  Thus, the duality transformation $\text{CS}_{\mathcal{L}}$ yields a Hamiltonian of uncoupled sites, where each site Hamiltonian takes the form of a transverse-field Ising model on some connected graph.

With this mapping, the spectrum of the Hamiltonian~\eqref{hamilt} can be calculated explicitly, with the Hamiltonian term $H_\mu'$ at each site, for example by using a Jordan-Wigner transformation.  We note at this point that each Hamiltonian $H_\mu'$ has a nondegenerate ground state for all $\lambda>0$; thus, our PEPS Hamiltonian on the full lattice will also possess a nondegenerate ground state for $\lambda>0$.

\subsubsection{Encoded stabilizers:  Constants of motion}

For each site $\mu$, define the operator
\begin{equation}
  K_\mu:=\bigotimes_{i=1}^c\sigma^x_{(\mu,i)}\otimes\sigma^z_{(\nu(i),i)}\,.
\end{equation}
That is, $K_\mu$ is the tensor product of $\sigma^x$ for every qubit at site $\mu$ together with $\sigma^z$ on every neighbouring site.  It is straightforward to show that all such operators commute with the Hamiltonian of Eq.~\eqref{hamilt},
\begin{equation}
  [K_\mu, H]=0\,, \quad \forall\ \mu \in \mathcal{L}\,,
\end{equation}
and with each other,
\begin{equation}
  [K_\mu,K_\nu]=0\,, \quad \forall\ \mu,\nu\in\mathcal{L}\,.
\end{equation}
Thus, if $H$ has a nondegenerate ground state (as is the case for the PEPS Hamiltonian with $\lambda >0$), it must also be a simultaneous eigenstate of all operators $K_\mu$.

Using the duality transformation $\text{CS}_{\mathcal{L}}$ defined above, we find that
\begin{equation}
  (\text{CS}_{\mathcal{L}})K_\mu(\text{CS}_{\mathcal{L}}) = \bigotimes_{i=1}^c \sigma^x_{(\mu,i)} \,.
\end{equation}
Using the well-known solution to the transverse-field Ising model with Hamiltonian~(\ref{eq:TFIM}), we find that the ground state for $\lambda>0$ is the $+1$ eigenstate of this operator.  (This can be inferred by the fact that the ground state is clearly the $+1$ eigenstate of $\otimes_{i=1}^c\sigma^x_{(\mu,i)}$ in the limit $\lambda/g\rightarrow \infty$.)  Thus, we have that the ground state of the PEPS Hamiltonian for $\lambda>0$ is the simultaneous $+1$ eigenstate of all operators $K_\mu$, $\mu \in \mathcal{L}$.

The operators $K_\mu$, then, can be viewed as \emph{encoded cluster stabilizers}, and the ground state for $\lambda>0$ as an encoded cluster state.  Unfortunately, for $\lambda>0$, this encoding is no longer in the ground state space of $H_0$ spanned locally at sites by the states $|00\ldots 0\rangle$ and $|11\ldots 1\rangle$.  Our perspective is to consider the encoding to be fixed in this space and view the ground state of the PEPS Hamiltonian as the desired locally-encoded cluster state plus perturbative corrections in $\lambda/g$.  These concepts are best illustrated through a simple example.

\section{Example: a 1-D Line} \label{PertApp}

First we illustrate this approach on the simplest lattice: a one-dimensional line of qubits with periodic boundary conditions.  We demonstrate explicitly that the perturbative procedure yields a low-energy effective Hamiltonian on the logical qubits of the form of Eq.~(\ref{eq:ClusterHamiltonian}), and that an approximate cluster state is obtained as the non-degenerate ground state.  The basic steps outlined in this example for the one-dimensional line, suitably generalized, will be applicable to more complex lattice structures.

%\section{Lattice Structure}\label{1D line}

\begin{figure}[!]
\begin{center}
\includegraphics[width=80mm]{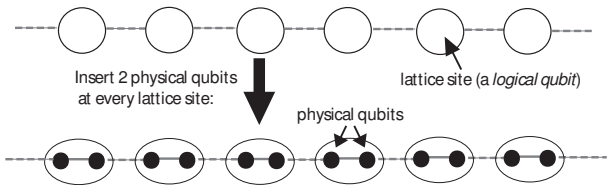}
   \caption{The logical lattice and the physical lattice structure for the 1D line.}\label{fig:1Dline}
\end{center}
\end{figure}

Consider a one-dimensional line consisting of $N_S$ qubit sites with periodic boundary conditions (i.e., a ring).  The coordination number of this lattice is $c=2$, and thus our construction requires two qubits to be placed at each site.  This lattice structure is illustrated in Fig.~\ref{fig:1Dline}.  The Hamiltonian is that of Eq.~(\ref{hamilt}).

%\section{The Form of the Hamiltonian}
\subsection{The unperturbed spectrum}

We first investigate the energy eigenvalue spectrum of the
unperturbed Hamiltonian $gH_0$.  Because $gH_0$ is a sum of terms, each of the form $\sigma^z \otimes \sigma^z$ acting on a single site, the energy spectrum can be determined by
analysing each site individually.  At a single site, there are two
energy levels. The ground-state is degenerate, two-dimensional, and
spanned by the states
\begin{equation}
  |00\rangle =: |\mathbf{0}_L\rangle\,, \quad
  |11\rangle =: |\mathbf{1}_L\rangle\,.
\end{equation}
The ground state space of the unperturbed Hamiltonian at each site,
then, can be viewed as a \emph{logical qubit}. Note also that this
ground state space is, by construction, the logical subspace for the
cluster state PEPS projection. The energy
of this ground state space is $-g$. The first excited state at each site is also
two-dimensional, has an energy of $g$, and is spanned by the states
$|01\rangle$ and $|10\rangle$.

With the spectrum of $gH_0$ at each site, we now describe
the spectrum of the unperturbed Hamiltonian $gH_0$ on the entire
lattice.  The lattice ground-state space is spanned by product states of
all of the individual sites in the ground state (i.e. in the
logical space). This ground-state space has energy $E_0^{(0)} =
-gN_S$, is $2^{N_S}$-dimensional, and is spanned by all logical
states of $N_S$ qubits.  We denote this space $\mathcal{H}_L$.
The first-excited space is $(2N_S\cdot 2^{N_S-1})$-dimensional, and
has energy $E_1^{(0)} = -g(N_S-2)$.  Thus, for the unperturbed
Hamiltonian $gH_0$, the gap from the ground to first-excited space
is $2g$.  The second-excited space has energy $E_2^{(0)} =
-g(N_S-4)$, and so on. These energies will serve as the zeroth-order energies
in perturbation theory for the total Hamiltonian.

%\section{Applying the Perturbation: Breaking the Degeneracy}
\subsection{Perturbation theory}

We now turn to perturbation theory and determine the effect of the
term $\lambda V$ in the Hamiltonian~(\ref{hamilt}).  We will show that this term lifts
degeneracy of the ground state, and that the logical cluster state arises as the unique ground state (although we also show that there are perturbative corrections to this state).  For details of our use of perturbation theory and notation, see the Appendix.

Let the $n$th-order energy
correction to the $j$th state in $\mathcal{H}_L$ be denoted by
$\lambda^n E_{0j}^{(n)}$.  Let $P_L$ be the projection onto the
degenerate ground state space of the unperturbed Hamiltonian $gH_0$,
i.e., onto the ``logical'' space $\mathcal{H}_L$.  Define $\bar{P}_L:=I-P_L$ to be the projection onto the
``illogical space'' (denoted $\mathcal{H}_{\bar{L}}$) and let the projection onto the $j$th
unperturbed excited level be denoted $P_j$.

\begin{figure}[!]
\begin{center}
\includegraphics[width=60mm]{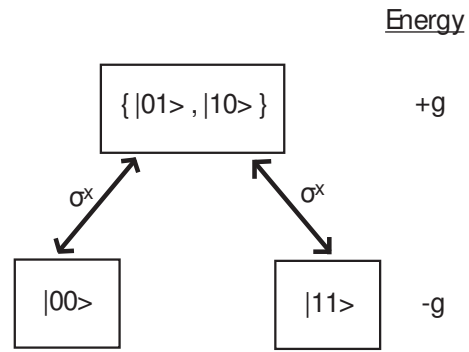}
   \caption{The effect of the $\sigma^x$ terms in $V$ on a single site.}\label{fig:1Denergy}
\end{center}
\end{figure}

To obtain a conceptual view of the perturbation it is useful to see
the effect of $V$ on a single site. The Hamiltonian $V$ is a sum of terms of the
form $-\sigma^z\otimes \sigma^x$; however, each of the Pauli operators in such a term act on different sites, and so we must consider the action of $\sigma^x$ and $\sigma^z$ separately.  Because the logical space
$\mathcal{H}_L$ is spanned by $|00\rangle$ and $|11\rangle$, the action of $\sigma^x$ will move a
site out of the logical space; the action of $\sigma^z$ will not, and simply induce a phase.
The possible actions by the $\sigma^x$ part of $V$ at a single site are shown in Fig.~\ref{fig:1Denergy}.

The first-order corrections to the energy are governed by the operator (see Eq.~(\ref{A11}) in the Appendix)
\begin{equation}
  \label{1stord}
  \theta^{(1)}=P_L V P_L \,.
\end{equation}
Specifically, the first order energy corrections $E_{0j}^{(1)}$ to the ground
state are the eigenvalues of this operator.  Because all of the terms in $V$ contain a single $\sigma^x$, they all move a
state in the logical space to the first excited state (i.e., $ V P_L
= P_1 V P_L$).  Thus, $P_L V P_L = 0$, and there is no first-order correction to the
energies.

The second-order corrections are governed by the operator
\begin{align}
  \label{2ndord}
  \theta^{(2)} &= P_L V \bar{P}_L
  (E_0^{(0)}-gH_0)^{-1} \bar{P}_L V  P_L \nonumber \\
  &= \frac{P_L V P_1 V
  P_L}{(E_0^{(0)}-E_1^{(0)})} \,.
\end{align}
where the expression has been simplified using $ V P_L = P_1 V P_L$.
The operator $ P_L V P_1 V P_L $ maps states from the ground state space to the first excited space and then back to ground state space.
By investigating the different ways of
returning to the logical space after just two $\sigma^x$ operations,
it is clear that there are two possible contributions to this term:
\begin{enumerate}
\item If $\sigma^z\otimes\sigma^x$ in $V$ is applied twice to the same bond,
it yields the identity. The first $\sigma^x$ can be applied to any of the qubits
and then must be applied again to the same qubit, so there are $2N_S$ of these terms.
\item If $\sigma^x$ is applied to each of the two qubits at a
site (and corresponding $\sigma^z$ operations to qubits in the neighboring
sites), then the lattice remains in the ground state. We can apply
the first $\sigma^x$ to either of the two qubits at the site and so
there will be two of these terms that occur at each site. Explicitly,
this case will be a term applied to the logical space of the form
\begin{equation}
  K_\mu:=\bigotimes_{i=1,2}\sigma^x_{(\mu,i)}\otimes\sigma^z_{(\nu(i),i)}\,,
\end{equation}
where $\mu \sim \nu$.

The operator $S_\mu := P_L K_\mu P_L$, which
acts only on the logical space, can be determined
explicitly as follows. Note that the product of two $\sigma^x$
operators on a single site $\mu$ (one on each physical qubit), restricted to the logical space, is
equivalent to a \emph{logical} $X$ operator
\begin{equation}
  X_{\mu} :=  P_L
  \sigma^x_{(\mu,1)}\sigma^x_{(\mu,2)}P_L \,.
\end{equation}
Also, a single $\sigma^z$ operator acting on either of the two
physical qubits at a site $\nu$, restricted to the logical space, is equivalent to a \emph{logical}
$Z$ operator
\begin{equation}
  Z_{\nu} := P_L \sigma^z_{(\nu,i)}P_L\,.
\end{equation}
Thus, $S_\mu = X_\mu \otimes_{\nu\sim\mu}Z_\nu$.  This operator is a (logical) stabilizer of the cluster state on this lattice.
\end{enumerate}
Therefore, we have
\begin{equation}
  P_L V P_1 V P_L =  2N_S P_L + 2 \sum_{\mu \in \mathcal{L}} S_\mu \,.
\end{equation}
Substituting this result into Eq.~(\ref{2ndord}) and using
$E_0^{(0)}-E_1^{(0)}=-2g$ gives
\begin{equation}
\label{2ndord2}
 \theta^{(2)}=\frac{2N_S P_L + 2 \sum_{\mu \in \mathcal{L}}
  S_\mu }{-2g} \,.
\end{equation}
The energies $E_{0j}^{(2)}$ are the eigenvalues of $\theta^{(2)}$ and the
corresponding eigenstates of $\theta^{(2)}$ will be the zeroth-order
energy eigenstates after the degeneracy is lifted.

Next, we identify the basis which diagonalises $\theta^{(2)}$; this is straightforward given the expression (\ref{2ndord2}).  As the cluster state is the simultaneous $+1$ eigenstate of all stabilizer operators $S_\mu$, the \emph{logical cluster state} on this lattice, denoted by $|C\rangle$, is an eigenstate of $\theta^{(2)}$.  Similarly, the other eigenstates of $\theta^{(2)}$ are also just the simultaneous eigenstates of the stabilizers $S_\mu$ (as all such stabilizers commute).  Explicitly, let $|C\{\alpha,\beta,\ldots\}\rangle$ denote the logical cluster state with a logical $Z$-operator (called a \emph{$Z$-error}) applied to
the sites $\alpha,\beta,\ldots \in \mathcal{L}$.  Using the anti-commutation relations of the Pauli matrices,
$|C\{\alpha,\beta,\ldots\}\rangle$ is the $-1$ eigenstate of $S_\alpha,S_\beta,...$ and the $+1$ eigenstate of $S_\mu$ for $\mu \neq \alpha,\beta,\ldots$.  Therefore, the $2^{N_S}$ states of the form $|C\{\alpha,\beta,\ldots\}\rangle$ will be eigenstates of $\theta^{(2)}$. Furthermore, these states are orthogonal, as each pair of states will have a differing
eigenvalue for least one of the $S_\mu$ operators. In summary, the set of states
$\{|C\rangle,|C\{\alpha\}\rangle,|C\{\alpha,\beta\}\rangle,\ldots \}$, running over logical $Z$-errors at all possible sites, forms an
orthogonal basis of $\mathcal{H}_L$ and diagonalises $\theta^{(2)}$.

The eigenvalue spectrum of $\theta^{(2)}$ is then straightforward to
calculate using the properties of stabilisers.  From the form of
$\theta^{(2)}$ in Eq.~(\ref{2ndord2}), the lowest energy
eigenstate will be the cluster state $|C\rangle$, because it is an
eigenstate of all stabilisers in the sum $\sum_{\mu \in \mathcal{L}} S_\mu $
with eigenvalue $+1$. Thus the second-order correction for the
energy associated with the cluster state is
\begin{equation}
    \lambda^2 E^{(2)}_{|C\rangle}
    = \frac{2N_S+2N_S}{-2g}\lambda^2
    = -2N_S\frac{\lambda^2}{g} \,.
\end{equation}
Next, consider a state $|C\{\alpha\}\rangle = Z_\alpha|C\rangle$, a
cluster state with a single $Z$-error at the site $\alpha$.  This
state is also an eigenstate of all stabilizers in the sum $\sum_{\mu
\in \mathcal{L}} S_\mu$ with eigenvalue $+1$ \emph{except} the stabilizer
$S_\alpha$ for which it has eigenvalue $-1$. Therefore,
\begin{equation}
    \lambda^2 E^{(2)}_{|C\{\alpha\}\rangle}
    = \frac{2N_S+2(N_S-2)}{-2g}\lambda^2
    = -2(N_S-1)\frac{\lambda^2}{g} \,.
\end{equation}
Because there are $N_S$ states of the form $|C\{\alpha\}\rangle$, this
$1^{\text{st}}$ excited space is $N_S$-dimensional. Similarly, the
$n$th excited space up to $n=N_S$ is
$\binom{N_S}{n}$-dimensional and (to zeroth order) is spanned by
states obtained from $|C\rangle$ by $n$ logical $Z$-errors.

Higher order corrections can be calculated by following a similar procedure.  As noted in Sec.~\ref{subsec:ExactSol}, this Hamiltonian can be easily solved exactly, with a ground state energy given by
\begin{equation} \label{1Dlinetot}
    E_{|C\rangle}
    = -gN_S \sqrt{1+4\frac{\lambda^2}{g^2}} \,.
\end{equation}
There is an energy gap
\begin{align}
  \Delta &:= g\Bigl( \sqrt{1+4\frac{\lambda^2}{g^2}}-1\Bigr) \nonumber \\
  & \simeq 2\lambda^2/g + O(\lambda^3/g^2) \,,
\end{align}
to the first excited space; all higher levels have energy $E_n = E_{|C\rangle} + n\Delta$.  Note that $\Delta$ is independent of $N_S$, the size of the lattice. Intuitively, then, one may associate logical $Z$ errors on any site with a fixed energy
$\Delta$ each.

In summary, we have shown that the non-degenerate ground state of the Hamiltonian $H=gH_0+\lambda V$ is the cluster state, to zeroth order in $\lambda/g$, with an energy gap to the second excited state scaling as $\sim\lambda^2/g$.

\subsection{Perturbative corrections to the ground state}

We have shown that, to zeroth order in $\lambda/g$, the
ground state of the system is the logical cluster state $|C\rangle$.
However, the perturbation will also modify the energy eigenstates
from their unperturbed states. To first order in $V$, the perturbed
ground state $|E_0\rangle$ is given (up to normalization) as
\begin{multline}
  |E_0\rangle \propto  |C\rangle + \lambda \Bigl(\sum_{|j\rangle \in
  \mathcal{H}_{\bar{L}}} \frac{ \langle j | V |C\rangle}{
  E_0^{(0)}-E_j^{(0)}}|j\rangle \\
  + \sum_{|l\rangle \in \mathcal{H}_L, |l\rangle
  \neq |C\rangle} \frac{\langle l |\theta^{(3)}
  |C\rangle}{E_{|C\rangle}^{(2)}-E_l^{(2)}}|l\rangle \Bigr)\,,
\end{multline}
where
\begin{equation}
  \theta^{(3)}= P_L V [(E_0^{(0)}-H_0)^{-1} \bar{P}_L V]^2 P_L\,.
\end{equation}
For this perturbation, $\theta^{(3)}=0$; however, there exist states $|j\rangle \in \mathcal{H}_{\bar{L}}$
such that $\langle j | V |C\rangle \neq 0$.

Note that $V$ is a sum of $2N_S$ terms of the form $\sigma^x \otimes \sigma^z$ acting across a bond.  Each of these terms applied to $|C\rangle$ gives an excited state of the form
\begin{equation}
  |k_{(\mu,i)}\rangle:=\sigma^x_{(\mu,i)}\otimes \sigma^z_{(\nu(i),i)}|C\rangle\,.
\end{equation}
Using the anti-commutation relations of the Pauli matrices, the terms in
$H_0$ act on $|k_{(\mu,i)}\rangle$ as
\begin{align}
  (\sigma^z_{(\mu,i)}\otimes
  \sigma^z_{(\mu,i+ 1)}) |k_{(\mu,i)}\rangle & =- |k_{(\mu,i)}\rangle \,, \\
  (\sigma^z_{(\rho,i)}\otimes \sigma^z_{(\rho,i+1)}) |k_{(\mu,i)}\rangle &= |k_{(\mu,i)}\rangle\,, \quad \rho \neq \mu\,.
  \label{operator1}
\end{align}
Hence $gH_0|k_{(\mu,i)}\rangle=
-g(N_S-2)|k_{(\mu,i)}\rangle=E_1^{(0)}|k_{(\mu,i)}\rangle$, and therefore the states $|k_{(\mu,i)}\rangle$ are in the first excited space
of $gH_0$.  Eq.~(\ref{operator1}) also shows that $|k_{(\mu_1,i_1)}\rangle$ and
$|k_{(\mu_2,i_2)}\rangle$ for $\mu_1 \neq \mu_2$ are eigenvectors
with different eigenvalues for the operator
$\sigma^z_{(\mu_1,1)}\otimes \sigma^z_{(\mu_1,2)}$ and thus they are
orthogonal.  However, recalling from earlier that $K_{\mu}:=
\otimes_{i=1}^2 \sigma^x_{(\mu,i)}\otimes \sigma^z_{(\nu(i),i)}$
stabilises $|C\rangle$, we have that $\langle
k_{(\mu,1)}|k_{(\mu,2)}\rangle= \langle C| K_{\mu} | C\rangle = 1$
and thus $|k_{(\mu,1)}\rangle=|k_{(\mu,2)}\rangle$. Hence $\langle
k_{(\mu,1)} | V |C\rangle=2$, and
\begin{align}\label{1Dstatecorr}
  |E_0\rangle &\propto  |C\rangle + \lambda\sum_{\mu \in \mathcal{L}} \frac{ \langle k_{(\mu,1)} | V |C\rangle}{ E_0^{(0)}-E_1^{(0)}}|k_{(\mu,1)}\rangle \notag\\
 &\propto  |C\rangle - \frac{\lambda}{g}\sum_{\mu \in \mathcal{L}} |k_{(\mu,1)}\rangle \,.
\end{align}
There are $N_S$ states in the above sum, which determines the normalization.  Thus, we can calculate the fidelity $F = |\langle C|E_0\rangle|^2$ of the ground state with the exact cluster state, which in this case is found to be
\begin{equation}
  F = \frac{1}{1+N_S \lambda^2/g^2} \,.
\end{equation}
This fidelity decays rapidly for increasing $N_S$, which is unsurprising given that it is comparing quantum states on a large lattice and is an extensive quantity.  For any lattice system with $N_S$ large, this fidelity is known to scale as $F = d^{N_S}$, where $d$ is an intensive quantity that can be interpreted as the average fidelity per site~\cite{Zho07}.  Precisely,
\begin{equation}
  \label{eq:avgfidelitypersite}
  \log d := \lim_{N_S\to\infty} N_S^{-1}\log F\,,
\end{equation}
which is found to satisfy
\begin{align}
  \log d &= -N_S^{-1}\log (1+N_S \lambda^2/g^2) \nonumber \\
  &> -N_S^{-1}\log (1+\lambda^2/g^2)^{N_S} \nonumber \\
  &= -\log(1+\lambda^2/g^2) \,.
\end{align}
Thus $d > (1+\lambda^2/g^2)^{-1}$, which is independent of $N_S$.  This result demonstrates that the ground state is ``close'' to the ideal cluster state, as quantified by a large average fidelity per site, for $\lambda\ll g$.

\section{Universal resources for MBQC}

Although it serves as an illustrative example of the techniques presented in this paper, the cluster state on a line is not a universal resource for MBQC; a higher-dimensional lattice is required.  In this section, we apply the perturbative procedure to lattice structures that are interesting from a MBQC perspective, and comment on their utility.

\subsection{Hexagonal lattice}

\begin{figure*}[!]
\begin{center}
\includegraphics[width=160mm]{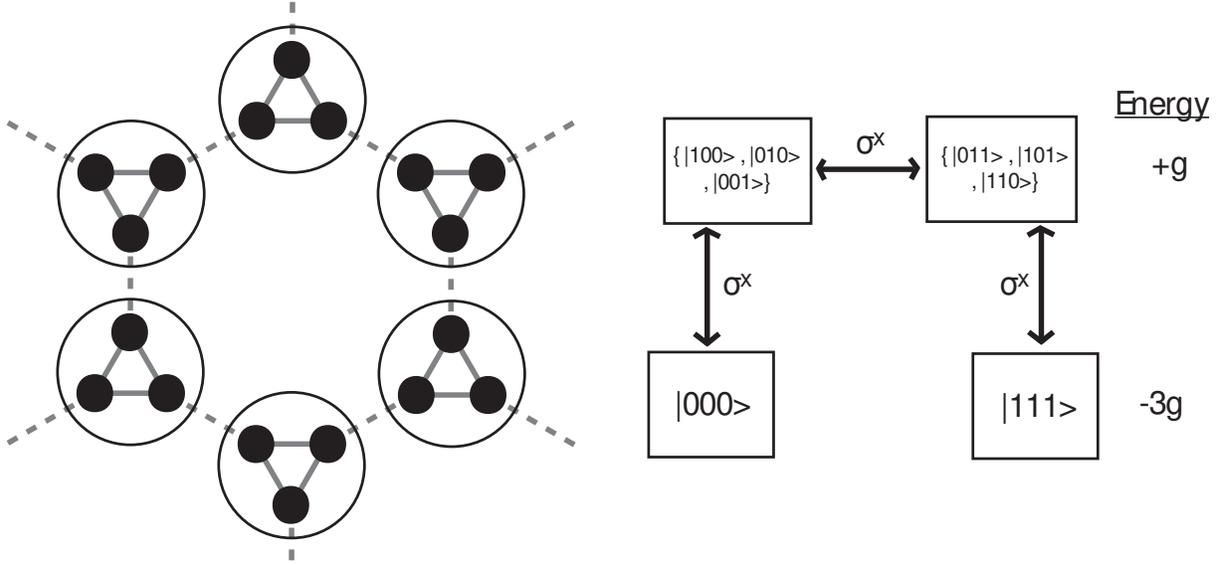}
   \caption{The hexagonal lattice structure, and the effect of the $\sigma^x$ terms in $V$ on a single site.}\label{fig:2Dhexenergy}
\end{center}
\end{figure*}

For a hexagonal lattice in two dimensions with $N_S$ sites and periodic boundary
conditions, the coordination number is 3, and we require three physical qubits at
each site (see Fig.~\ref{fig:2Dhexenergy}).  The Hamiltonian for
the lattice is again given by Eq.~(\ref{hamilt}).

We investigate the spectrum of the unperturbed Hamiltonian $gH_0$ by considering its action at a single site, where the three site qubits interact via the Ising coupling on a ring.  The ground-state is degenerate, two-dimensional, and spanned by the
states
\begin{equation}
  |000\rangle =: |\mathbf{0}_L\rangle\,, \quad
  |111\rangle =: |\mathbf{1}_L\rangle\,,
\end{equation}
which encode our logical qubit.  The energy of this
ground state space is $-3g$.  The first excited state is
six-dimensional, and has an energy of $g$.  Thus, for the entire
lattice of $N_S$ sites, the ground-state space has energy $E_0^{(0)}
= -3gN_S$, is $2^{N_S}$-dimensional, and is spanned by all logical
states of $N_S$ qubits. The first-excited space is $(6N_S\cdot
2^{N_S-1})$-dimensional, and has energy $E_1^{(0)} = -g(3N_S-4)$.

We now turn to perturbation theory.  It is again useful to obtain a
conceptual view of the effect of $V$ on a single site, as illustrated in Fig.~\ref{fig:2Dhexenergy}.  As the ground state space is spanned by $|000\rangle$ and $|111\rangle$ at each site, only the action of $\sigma^x$ (and not $\sigma^z$) will map states out of the logical ground state space. The possible actions by the $\sigma^x$ part of $V$ at a
single site are shown in Fig.~\ref{fig:2Dhexenergy}.
Once again, $P_L V P_L = 0$ and
there is no first-order correction to the energies.

The second order corrections $E_{0j}^{(2)}$ are the eigenvalues of
the operator $\theta^{(2)}$ defined in Eq.~(\ref{opert}). To evaluate the operator $ P_L V P_1 V P_L $, we examine Fig.~\ref{fig:2Dhexenergy} and the ways of
returning to the logical space after just two applications of $V$.  It is clear that there is only one possible contribution: if a $\sigma^z\otimes\sigma^x$ in $V$ is applied twice to the same bond, this will yield the identity.  The first $\sigma^x$ can be
applied to any of the qubits and then must be applied again to the
same qubit, so there are $3N_S$ such terms. Hence
\begin{equation}
  P_L V P_1 V P_L =  3N_S P_L \,.
\end{equation}
Using this result in Eq.~(\ref{2ndord2}) as well as $E_0^{(0)}-E_1^{(0)}=-4g$ gives
\begin{equation}
 \label{2ndord3}
 \theta^{(2)}=\frac{3N_S P_L }{(-4g)} \,.
\end{equation}
This operator simply acts as the identity on the logical space and
so there is a constant second-order correction to the ground-state
energy -- an energy shift -- given by
\begin{equation}
    \lambda^2 E_0^{(2)}
    = -\frac{3N_S\lambda^2}{4g} \,.
\end{equation}
The energy degeneracy of the ground state has still not been
broken at second order and we must proceed to third order.

The third order corrections $E_{0j}^{(3)}$ are the eigenvalues of
the operator $\theta^{(3)}$ given by
\begin{align}
  \theta^{(3)} &= P_L V \bigl[(E_0^{(0)}-H_0)^{-1} \bar{P}_L V \bigr]^2 P_L \nonumber \\
  &= \frac{P_L V P_1 V P_1 V P_L}{(E_0^{(0)}-E_1^{(0)})^2}\,,
\end{align}
where the expression has been simplified using $ V P_L = P_1 V P_L$.
With three applications of the perturbation $V$, the operator $ P_L
V P_1 V P_1 V P_L $ maps states out of the ground state space and then back again via the first excited
space.  Again investigating Fig.~\ref{fig:2Dhexenergy},
it is only possible for the lattice to remain in the ground state
after three perturbation terms if $\sigma^x$ operators are applied to each of the three qubits at a site (and, through $V$, the corresponding $\sigma^z$ operators to the qubits on each of the neighboring sites).  That is, this case
will be a term applied to the logical space of the form
\begin{equation}
  K_\mu:=\bigotimes_{i=1,2,3}\sigma^x_{(\mu,i)}\otimes\sigma^z_{(\nu(i),i)}\,,
\end{equation}
where $\nu(i)$ is the site connected to $(\mu,i)$ by a bond.  Just as in the case of the line,
the operator $K_\mu$ acts on the logical space as $S_\mu := P_L K_\mu P_L = X_\mu
\prod_{\nu\sim\mu}Z_\nu$, a logical cluster-state stabilizer operator. The three qubits at the site can be
ordered in $3!$ possible ways, and so there will be $3!$ of these terms that occur at each site.
Therefore,
\begin{equation}
  P_L V P_1 V P_1 V P_L =-3!\sum_{\mu \in \mathcal{L}} S_\mu\,,
\end{equation}
and
\begin{equation}
  \theta^{(3)}=  \frac{-3! \sum_{\mu \in \mathcal{L}} S_\mu}{(-4g)^2}=
  -\frac{3}{8g^2} \sum_{\mu \in \mathcal{L}} S_\mu \,.
\end{equation}

Once again, as in the line, the set of $2^{N_S}$ states
$\{|C\rangle,|C\{\alpha\}\rangle,|C\{\alpha,\beta\}\rangle,\ldots
\}$, running over logical $Z$-errors on the cluster state $|C\rangle$ at all possible sites, forms an
orthogonal basis of $\mathcal{H}_L$ and diagonalises $\theta^{(3)}$.
The cluster state $|C\rangle$ is the unique lowest eigenstate of
$\theta^{(3)}$.  The third-order
correction for the energy associated with this state is
\begin{equation}
    \lambda^3 E^{(3)}_{|C\rangle}
    = -\frac{3}{8}N_S\frac{\lambda^3}{g^2} \,.
\end{equation}
Again, this case is simple enough to analyze analytically; the ground state of the Hamiltonian $H=gH_0+\lambda V$ has energy
\begin{align}
    E_{|C\rangle} &= -gN_S \Bigl( 1+\frac{\lambda}{g}+2\sqrt{\frac{\lambda^2}{g^2}+\frac{\lambda}{g}+1}\Bigr) \nonumber \\
    &\simeq -3gN_S\Bigl(1 +\frac{1}{4}\frac{\lambda^2}{g^2} +\frac{1}{8}\frac{\lambda^3}{g^3} \Bigr)\,.
\end{align}
The $n$th excited space up to $n=N_S$ is
$\binom{N_S}{n}$-dimensional and is spanned (to zeroth order) by
states obtained from $|C\rangle$ with $n$ logical $Z$ errors. These
states have energy $E_n = E_{|C\rangle} + n\Delta$ where
\begin{align}
  \Delta &:= 2\Bigl(\frac{\lambda}{g} -\sqrt{\frac{\lambda^2}{g^2} + \frac{\lambda}{g} + 1} +\sqrt{\frac{\lambda^2}{g^2} - \frac{\lambda}{g} + 1}\Bigr) \nonumber \\
  &\simeq \frac{3}{4}\frac{\lambda^3}{g^2} + O(\lambda^4/g^3) \,.
  \label{3gap}
\end{align}

We can also calculate the first-order corrections to the ground state $|C\rangle$, by finding states $|j\rangle
\in \mathcal{H}_{\bar{L}}$ such that $\langle j | V |C\rangle \neq 0$.  As before, define
$|k_{(\mu,i)}\rangle:=\sigma^x_{(\mu,i)}\otimes \sigma^z_{(\nu(i),i)}|C\rangle$.  By determining the effect of each of the terms in $gH_0$ on $|k_{(\mu,i)}\rangle$ it is clear that they are in the first excited
space of $gH_0$ and are orthogonal to each other. To first order in $\lambda/g$,
\begin{equation}\label{corr3}
  |E_0\rangle \propto |C\rangle - \frac{\lambda}{4g}\sum_{\mu \in \mathcal{L}}
  \sum_{i=1}^3 |k_{(\mu,i)}\rangle \,.
\end{equation}
Comparing this ground state with the ideal cluster state, we find that the average fidelity per site $d$ is bounded by $d > (1+3\lambda^2/(4g)^2)^{-1}$.

\subsection{Square lattice}\label{2Dsq}

\begin{figure*}[!]
\begin{center}
\includegraphics[width=160mm]{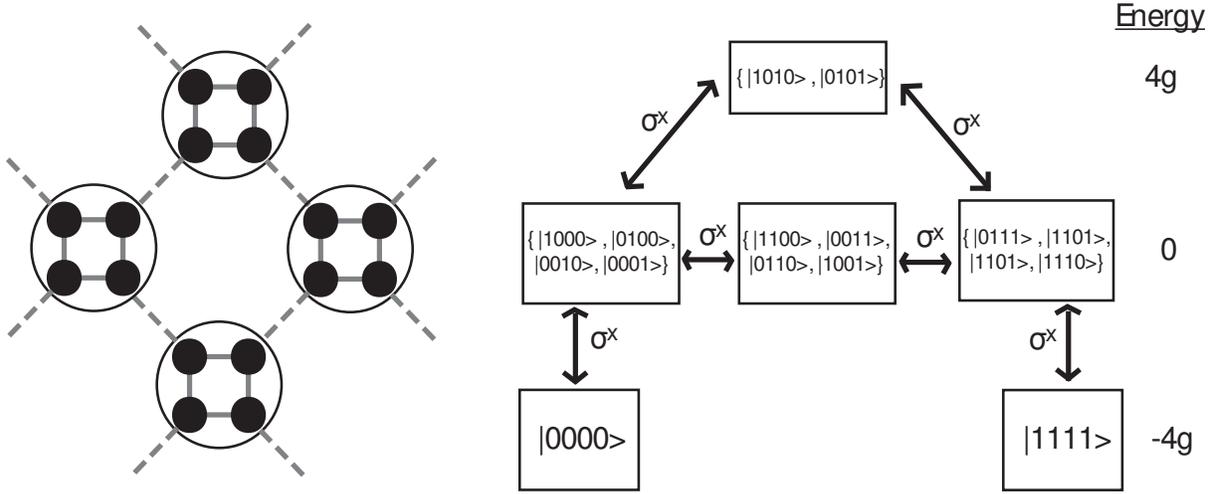}
   \caption{The square lattice structure, and the effect of the $\sigma^x$ terms in $V$ on a single site.}\label{fig:2Dsquareenergy}
\end{center}
\end{figure*}

We now repeat the above procedure for a 2D square lattice with $N_S$
sites and periodic boundary conditions.  This case was originally
examined in~\cite{Bar06}; however, our detailed derivation reveals some errors in their calculation of the perturbed energies and the gap.

The coordination number of this lattice is $4$, and so four physical qubits are
necessary at each site (see Fig.~\ref{fig:2Dsquareenergy}). The
Hamiltonian for the lattice is again given by Eq.~(\ref{hamilt}), again with a ring of four qubits coupled via an Ising interaction.  There are now three energy levels of $gH_0$ at a
single site. The ground-state space of $gH_0$ at a single site is spanned by the
states
\begin{equation}
  |0000\rangle =: |\mathbf{0}_L\rangle\,, \quad
  |1111\rangle =: |\mathbf{1}_L\rangle\,,
\end{equation}
and the energy of this ground state space is $-4g$. The first
excited state is twelve-dimensional, and has an energy of $0$. The
second excited state is two-dimensional and has a energy of $4g$.
So, for the entire lattice of $N_S$ sites, the ground-state space
has energy $E_0^{(0)} = -4gN_S$, is $2^{N_S}$-dimensional, and is
spanned by all logical states of $N_S$ qubits. The first-excited
space has energy $E_1^{(0)} = -4g(N_S-1)$ and the
second-excited space has energy $E_2^{(0)} = -4g(N_S-2)$.

The possible actions by the $\sigma^x$ part of $V$ at a single site are shown in Fig.~\ref{fig:2Dsquareenergy}.  We now follow the identical procedure as done previously, and find
\begin{align}
  \theta^{(1)} &= 0 \,, \\
  \theta^{(2)} &= \frac{4N_S P_L }{(-4g)} \,, \\
  \theta^{(3)} &= 0 \,, \\
  \theta^{(4)} &= -\frac{5}{16g^3}\sum_{\mu \in \mathcal{L}} S_\mu -
  \frac{N_S}{16g^3} P_L \,. \label{theta4}
\end{align}
That is, there are no first- or third-order corrections to the energy; at second-order there is a constant energy shift $\lambda^2 E_0^{(2)} = -N_S\lambda^2/g$ to the ground state; at fourth-order the degeneracy is broken.  In the expression for $\theta^{(4)}$, the first term is recognized as proportional to the cluster Hamiltonian: the sum of stabilisers of the
cluster state $S_\mu$. Therefore the set of $2^{N_S}$ states $\{|C\rangle,|C\{\alpha\}\rangle,|C\{\alpha,\beta\}\rangle,\ldots
\}$, running over logical $Z$-errors at all possible sites on the cluster state $|C\rangle$, is an
orthogonal basis of $\mathcal{H}_L$ which diagonalises $\theta^{(4)}$.

The cluster state $|C\rangle$ is the unique lowest eigenstate of
$\theta^{(4)}$, because it is an eigenstate of all stabilisers in the
sum $\sum_{\mu \in \mathcal{L}} S_\mu$ with eigenvalue $+1$. The fourth-order
correction for the energy associated with this state is
\begin{equation}
    \lambda^4 E^{(4)}_{|C\rangle} = -\frac{3}{8}N_S\frac{\lambda^4}{g^3} \,.
\end{equation}
We note that this result differs, by numerical factors, from the result of~\cite{Bar06}.  (The error in~\cite{Bar06} arises from missing contributions to the perturbation operator $\theta^{(4)}$ in Eq.~(\ref{theta4}).)  Higher-order corrections follow in a similar fashion, and a complete analytic solution for the ground state energy is found to be
\begin{align}
    E_{|C\rangle} &= -2gN_S\sqrt{2+2\frac{\lambda^2}{g^2} + 2\sqrt{\frac{\lambda^4}{g^4}+1}} \nonumber \\
    &\simeq -4gN_S\Bigl( 1 +\frac{1}{4}\frac{\lambda^2}{g^2} +\frac{3}{32}\frac{\lambda^4}{g^4} \Bigr)\,.
\end{align}

The $n$th excited space up to $n=N_S$ is
$\binom{N_S}{n}$-dimensional and is spanned (to zeroth order) by
states obtained from $|C\rangle$ by $n$ logical $Z$ errors. These
states have energy $E_n = E_{|C\rangle} + n\Delta$, where
\begin{align}
  \Delta &:= -2g\Bigl(1+\sqrt{\frac{\lambda^2}{g^2}+1} + \sqrt{2+2\frac{\lambda^2}{g^2} + 2\sqrt{\frac{\lambda^4}{g^4}+1}}\Bigr) \nonumber \\
  &\simeq \frac{5}{8}\frac{\lambda^4}{g^3} \,.
  \label{4gap}
\end{align}
Once again we can also calculate the first-order corrections to the
ground state $|C\rangle$ is calculated to be
\begin{equation}\label{corr4}
  |E_0\rangle \propto |C\rangle - \frac{\lambda}{4g}\sum_{\mu \in \mathcal{L}}
  \sum_{i=1}^4 |k_{(\mu,i)}\rangle \,,
\end{equation}
where $|k_{(\mu,i)}\rangle:=\sigma^x_{(\mu,i)}\otimes
\sigma^z_{(\nu(i),i)}|C\rangle$.  Comparing this ground state with the ideal cluster state, we again find that the average fidelity per site $d$ is bounded by $d > (1+4\lambda^2/(4g)^2)^{-1}$.

\subsection{Cubic lattice} \label{cubic}

\begin{figure*}[!]
\begin{center}
\includegraphics[width=160mm]{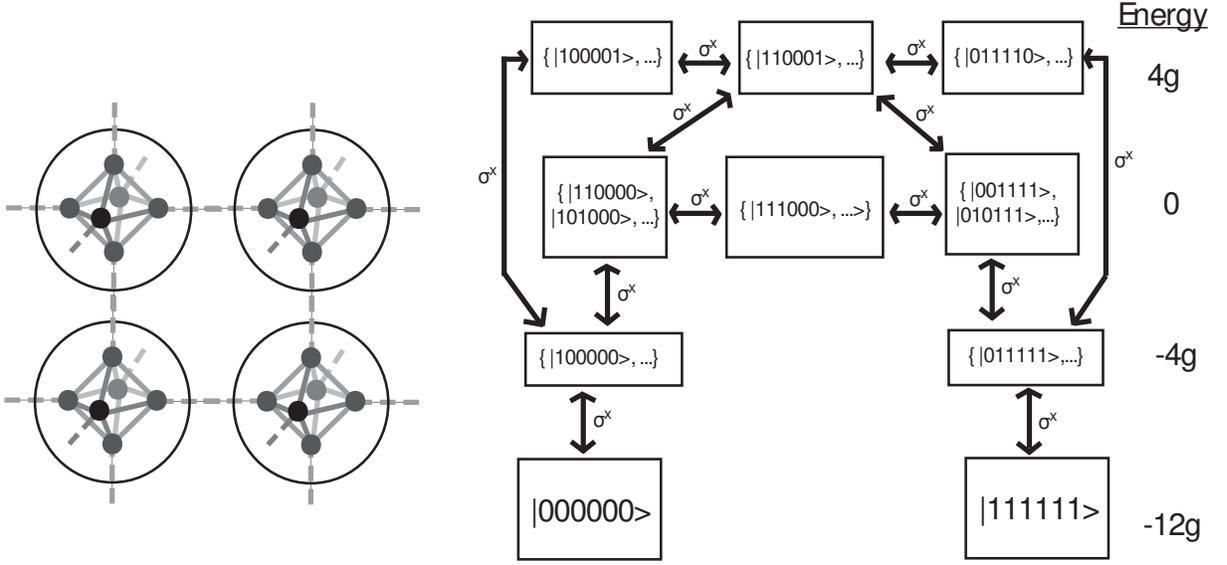}
   \caption{The cubic lattice structure, and the effect of the $\sigma^x$ terms in $V$ on a single site.}\label{fig:3Dcubeenergy}
\end{center}
\end{figure*}

We apply the now familiar procedure to a cubic lattice in three dimensions with
$N_S$ sites and periodic boundary conditions.  The coordination number is 6, and so six physical qubits are
necessary at each site (see Fig.~\ref{fig:3Dcubeenergy}). The
Hamiltonian for the lattice is given in Eq.~(\ref{hamilt}), where we arrange the $6$ qubits on the vertices of a octahedron, and place $\sigma^z \otimes \sigma^z$ couplings between all qubits connected by an edge of the octahedron, as in Fig.~\ref{fig:3Dcubeenergy}. There are four energy levels of $gH_0$ at a single
site; the ground-state is degenerate, two-dimensional, and spanned
by the states
\begin{equation}
  |000000\rangle =: |\mathbf{0}_L\rangle\,, \quad
  |111111\rangle =: |\mathbf{1}_L\rangle\,.
\end{equation}
The energy of this ground state space is $-12g$;  the first excited
state space has energy $-4g$; the second excited state space has energy $0$; the
third excited state space has energy $4g$.

The possible actions of $\sigma^x$ at a single site are shown in Fig.~\ref{fig:3Dcubeenergy}.  Again following our general perturbative procedure we find $\theta^{(1)} = \theta^{(3)} = \theta^{(5)} = 0$ and
\begin{align}
 \theta^{(2)} &= -\frac{6N_S}{8g}P_L \,, \\
 \theta^{(4)} &= -\frac{N_S}{256g^3}P_L \,, \\
 \theta^{(6)} &= -\frac{13N_S}{49152g^5}P_L-\frac{83}{16384g^5}\sum_{\mu \in \mathcal{L}} S_\mu \,.
\end{align}
Thus, there are two constant energy shifts at second and fourth order of
\begin{equation}
    \lambda^2 E_0^{(2)}
    = -\frac{3N_S\lambda^2}{4g} \,, \qquad \lambda^4 E_0^{(4)}
    = -\frac{N_S\lambda^4}{256g^3} \,.
\end{equation}
In the expression for $\theta^{(6)}$, the degeneracy is broken by the terms $S_\mu=X_\mu \otimes_{\nu\sim\mu}Z_\nu$ which are the cluster stabilizers. The set of $2^{N_S}$ states $\{|C\rangle,|C\{\alpha\}\rangle,|C\{\alpha,\beta\}\rangle,\ldots
\}$, running over logical $Z$-errors at all possible sites of the cluster state $|C\rangle$, forms an
orthogonal basis of $\mathcal{H}_L$ which diagonalises $\theta^{(6)}$.

The cluster state $|C\rangle$ is the unique lowest eigenstate of
$\theta^{(6)}$. The sixth-order correction for the energy associated with this state is
\begin{equation}
    \lambda^6 E^{(6)}_{|C\rangle}
    = -\frac{131}{24576}N_S\frac{\lambda^6}{g^5} \,.
\end{equation}
Therefore, to sixth order, the energy of the ground state is
\begin{equation}
    E_{|C\rangle}
    = -12gN_S \Bigl(1+\frac{1}{16}\frac{\lambda^2}{g^2} +\frac{1}{3\cdot 2^{10}}\frac{\lambda^4}{g^4}+\frac{131}{3^2\cdot 2^{15}}\frac{\lambda^6}{g^6} \Bigr) \,.
\end{equation}
The $n$th excited space up to $n=N_S$ is
$\binom{N_S}{n}$-dimensional and is spanned (to zeroth order) by
states obtained from $|C\rangle$ by $n$ logical $Z$ errors. These
states have energy $E_n = E_{|C\rangle} + n\Delta$, where
\begin{equation}\label{6gap}
  \Delta := \frac{83}{8192}\frac{\lambda^6}{g^5} \,.
\end{equation}
Once again we can also calculate the first-order corrections to the
ground state $|C\rangle$ is calculated to be
\begin{align}\label{corr6}
|E_0\rangle &= |C\rangle - \frac{\lambda}{8g}\sum_{\mu \in \mathcal{L}}
\sum_{i=1}^6 |k_{(\mu,i)}\rangle \,,
\end{align}
where $|k_{(\mu,i)}\rangle:=\sigma^x_{(\mu,i)}\otimes
\sigma^z_{(\nu(i),i)}|C\rangle$.   Comparing this ground state with the ideal cluster state, we find that the average fidelity per site $d$ is bounded by $d > (1+6\lambda^2/(8g)^2)^{-1}$.

\subsection{Implications for MBQC}\label{useful}

The cluster states on the three lattice types examined in this section (the
square, hexagonal and cubic lattices) are all universal resources
for quantum computation.  In each case, it has been shown above that
the perturbative procedure produces a non-degenerate ground state
which approximates an encoded cluster state on the lattice.  We chose to investigate each of these lattice structures because each has a unique relevance to the study of MBQC.  The 2D square lattice is the canonical example for use in
cluster-state quantum computing and was the original lattice structure presented in~\cite{Rau1}.  This lattice is also the most easily accessible to experimental investigation in cold atomic systems~\cite{Tre06}.  A hexagonal lattice was also examined above
because (as argued in~\cite{Bar06}) the perturbative procedure
produces a cluster state with the largest energy gap for a given ratio $\lambda/g$.  We discuss the implications of this observation below.  Finally, recent work~\cite{Rau06,Rau07} has
shown that fault-tolerant thresholds can be found for MBQC if the lattice used is
$3$-dimensional.

Following on from the discussion in~\cite{Bar06}, we now compare the
results for each lattice and relate it to its usefulness for quantum
computation. There are two sources of error when using the
ground state obtained in the perturbative procedure for
cluster-state quantum computation.

First, note that errors will arise because the ground state of the
system is not exactly the cluster state, but contains perturbative corrections (cf.~Eq.~(\ref{corr3}),
(\ref{corr4}), (\ref{corr6})). In each case the ground state is given by a superposition of the
cluster state with other first-excited states corresponding to ``errors'' $\sigma^x\otimes
\sigma^z$ applied to all bonds on the cluster state
independently.  This error rate is quantified by the average fidelity per lattice site $d$, defined by Eq.~\eqref{eq:avgfidelitypersite}, which was explicitly bounded in all of the above examples.  This bound takes the general form
\begin{equation}
  d > \frac{1}{1+k \lambda^2/g^2}\,,
\end{equation}
where $k$ is a constant of order one which depends on the lattice.  This bound tells us that, for $\lambda\ll g$, the ground state is very close to the cluster state, and that the error rate for the independent $\sigma^x\otimes
\sigma^z$ is less than $k\lambda^2/g^2$. Because we require $\lambda\ll g$, this error probability will be small.  The effect, and possible error correction, for such two-qubit correlated errors has not yet been investigated, but the independence and locality of the errors makes them amenable to existing error correction techniques.  In particular, we note that such an error can be identified by checking sites, each of which \emph{should} be in the code space spanned by $|00\ldots 0\rangle$ and $|11\ldots 1\rangle$.  Errors of the form $\sigma^x\otimes \sigma^z$ will cause a correctable error to this code space (which must also include a phase correction to the appropriate neighbouring site) provided that the lattice has coordination number $c>1$, i.e., for lattices of higher dimension than the 1-D line.  This correction scheme would require measurements of multiple qubits, and it would be worthwhile to investigate whether such error correction could be performed using single-qubit measurements.

The main difference arising in the calculations for each lattice structure, however, is the order in the perturbation theory at
which the ground-state degeneracy is broken. This occurs at third order for the hexagonal lattice, fourth order for the square lattice
and sixth order for the cubic lattice.  In general, the order at which perturbation theory breaks the ground-state degeneracy is given by the coordination number of the lattice.  This result leads directly to a dependence of the energy gap $\Delta$ on the coordination number $c$ of the lattice, as
\begin{equation}
  \Delta \sim ( \lambda/g)^c \,.
\end{equation}
In all cases the energy gap $\Delta$ is independent of the size of the lattice, i.e. the system is gapped.  Given that the rate at which the thermal state of this system will exhibit $Z$-errors depends explicitly on the size of this gap, the system will be
less sensitive to these errors if the energy gap $\Delta$ is made larger.  The hexagonal lattice will have the largest energy gap, as is consequently less sensitive to thermal errors.  It should be noted, however, that methods to identify and correct for such thermal errors (and Pauli errors in general) within the MBQC paradigm currently exist only for 3-dimensional lattices~\cite{Rau06,Rau07}.  (See also~\cite{Bar08}.)  The 2-D lattices (hexagonal and square) may not allow for error correction of such thermal errors using only single-qubit measurements; this remains a key open question.

We note that there exists a trade-off between these two types of errors when using the state for MBQC.  Increasing the value of $\lambda/g$ will reduce the probability of thermal errors at a given temperature but also perturb the ground state away from the
cluster state.

\section{Fixed Boundary Conditions}\label{fbc}

The perturbative approach has so far been successful
in producing the cluster state on each lattice type with
periodic boundary conditions. We now analyse the effect of placing fixed boundaries on the lattice.

\subsection{A line with fixed boundaries}

We first examine a line with fixed endpoints.  The interior sites still have coordination number $2$, and so we require two physical qubits at these sites. However, the boundary sites will consist of just a single physical qubit.  Denote the number of interior sites
by $N_S$, so that there are $(N_S+2)$ sites in the entire line. In
addition, denote the two boundary sites by the labels $\mu=B_1$ and
$\mu=B_2$.

The Hamiltonian for this lattice will remain that of Eq.~(\ref{hamilt}), where we do not place any site Hamiltonian term on the boundary sites.  The unperturbed energy spectrum at each of the interior sites is
unchanged from the periodic boundary case (as in Fig.~\ref{fig:1Denergy}).  The two boundary qubits, however, have zero
unperturbed energy.  The spectrum of the unperturbed
Hamiltonian for the entire line is therefore a $2^{N_S+2}$-dimensional
ground-state space with energy $E_0^{(0)} = -gN_S$, and is spanned
by all logical states of $(N_S+2)$ qubits.  The first-excited space
is $(2N_S\cdot 2^{N_S+1})$-dimensional, and has energy $E_1^{(0)} =
-g(N_S-2)$.

We now turn to perturbation theory.  Note that, for the two boundary qubits, a single application of $\sigma^x$ maps the logical space onto itself.  Thus, due to the contributions from the boundary qubits, there is now a first order correction to the
energy
\begin{align}
  \theta^{(1)} &= P_L V P_L \nonumber \\
  &= -P_L(K_ {B_1}+K_{B_2})P_L \nonumber \\
  &= -(S_ {B_1}+S_{B_2})\,,
\end{align}
where
\begin{align}
  K_{B_1}&= \sigma^x_{B_1} \otimes \sigma^z_{(\nu=1,1)}\,, \\
  K_{B_2}&= \sigma^x_{B_2} \otimes \sigma^z_{(\nu=N_S,2)} \,,
\end{align}
and as usual $S_\mu = X_\mu \otimes_{\nu\sim\mu}Z_\nu$.  In particular, the cluster stabilizers for the end sites are given by the product of an $X$ operator on the boundary site with a single $Z$ operator on its sole neighbour.

The first-order corrections to the ground-state energy,
$E_{0j}^{(1)}$, are the eigenvalues of $\theta^{(1)}$. This operator
is diagonal in our familiar basis for $\mathcal{H}_L$ of the
$2^{N_S+2}$ states
$\{|C\rangle,|C\{\alpha\}\rangle,|C\{\alpha,\beta\}\rangle,\ldots
\}$, running over logical $Z$-errors at all possible sites. The
states in this basis which are the $+1$-eigenstates of both $S_
{B1}$ and $S_{B2}$ will be the lowest eigenvalues of $\theta^{(1)}$.
By the property of stabilisers, the subspace $T \in \mathcal{H}_L$ which is stabilised by $S_ {B1}$ and $S_{B2}$ is
$2^{N_S}$-fold degenerate, and so the degeneracy is reduced by a factor of 4 at first order.  The cluster state $|C\rangle$ is
contained in this subspace, as are all states with logical $Z$-errors anywhere \emph{except} on the boundary.  Thus the lowest energy space $T$ has a first-order correction given by:
\begin{equation}
    \lambda E^{(1)}_{T}    = -2 \lambda \,.
\end{equation}
The next highest energy level includes states which have a $Z$-error
at either of the boundary sites $B_1$ or $B_2$ but not both. These
$2^{N_S+1}$ states have $\lambda E^{(1)}=0$, which is a gap of $2
\lambda$ above the space $T$.  The second highest energy level will
include states which have $Z$-errors at both boundary sites, and in
this case $\lambda E^{(1)}=2 \lambda$.

The second order correction is calculated in an identical manner to
the case with periodic boundary conditions. We have
\begin{equation}
 \theta^{(2)}=\frac{ P_L V  P_1 V  P_L}{(E_0^{(0)}-E_1^{(0)})} = \frac{2N_S  P_L + 2 \sum_{\mu \neq B_1, B_2} S_\mu}{-2g} \,.
\end{equation}
The above operator is already diagonal in our chosen basis (the
states of the form $|C\{\alpha,\beta,\ldots\}\rangle$). Of the
states in $T$, the cluster state $|C\rangle$ is the unique $+1$
eigenstate of all the stabilisers in the above sum, and so will be
the lowest eigenvalue of $\theta^{(2)}$. Thus the second-order
correction for the energy associated with this state is
\begin{equation}
    \lambda^2 E^{(2)}_{|C\rangle}
    = \frac{2N_S+2N_S}{-2g}\lambda^2
    = -2N_S\frac{\lambda^2}{g} \,.
\end{equation}
So in the case of fixed boundary conditions, the
cluster state is still the ground state produced (to zeroth order), with energy to 2nd order of
\begin{equation}
    E_{|C\rangle}
    = -gN_S-2\lambda -2N_S\frac{\lambda^2}{g} \,.
\end{equation}
A state $|C\{\alpha,\beta,\ldots\}\rangle$ with $n_B$ $Z$-errors at
boundary sites and $n_I$ $Z$-errors at interior sites will have
energy:
\begin{equation}
    E_{(n_B,n_I)}
    = E_{|C\rangle} + n_B \Delta_B + n_I \Delta_I
\end{equation}
where $\Delta_I := 2\lambda^2/g$ and $\Delta_B := 2\lambda$.  Provided $\lambda/g < 1$, the energy gap will be $\Delta_I
= 2\lambda^2/g$, the same energy gap which was obtained
with the periodic boundary conditions.

The first order corrections to this ground state will be given by
\begin{align}
  |E_0\rangle =  |C\rangle - \frac{\lambda}{g}\sum_{\mu \neq B_1, B_2} |k_{(\mu,1)}\rangle \,,
\end{align}
and the bound on $d$, the average fidelity per site, remains the same as for periodic boundary conditions.

\subsection{Square lattice with fixed boundaries}

Consider a square lattice of dimension $l \times l$.  We define the number of interior (non-boundary) sites to be $N_S$, and so $l = \sqrt{N_S}+2$ and the total number of sites is
$N_S+4\sqrt{N_S}+4$.  Interior sites have coordination number $4$, edge sites coordination number $3$, and corner sites coordination number $2$, determining the number of qubits at each site.  Denote the set of
corner boundary sites by $\mathcal{L}_1$, the set of edge boundary sites by $\mathcal{L}_2$, and the set of interior sites by $\mathcal{L}_3$. Each of
the three type of sites will have a different unperturbed spectrum at each, corresponding to Figs.~\ref{fig:1Denergy}, \ref{fig:2Dhexenergy} and \ref{fig:2Dsquareenergy} respectively. The zeroth order energies of
the lattice are now much more complicated due to the presence of
these three different types of sites. The ground state of $gH_0$ for
the entire lattice, spanned by all possible logical states, is
$2^{(N_S+4\sqrt{N_S}+4)}$-dimensional ground-state space with energy
$E_0^{(0)} = -4gN_S-12g\sqrt{N_S}-8g$. The next four excited states
separated by a energy gaps of $2g$.

At first-order in the perturbation, $\theta^{(1)} =
 P_L V  P_L = 0$, and thus there is still no first-order correction to
the energies.  At second, third, and fourth order, we have
\begin{align}
  \theta^{(2)} &= -\frac{(N_S+3\sqrt{N_S} +4)  P_L + \sum_{\mu \in \mathcal{L}_1} S_\mu}{g} \,, \notag \\
  \theta^{(3)}&= - \frac{3!\sum_{\mu \in \mathcal{L}_2} S_\mu}{(-4g)^{2}} \,, \notag \\
\theta^{(4)}&=- \frac{5}{16g^3}\sum_{\mu \in \mathcal{L}_3} S_\mu -
\frac{N_S+3\sqrt{N_S}-4+4n_1}{4g^3}\sum_{\mu \in \mathcal{L}_1} S_\mu
  \notag\\
&- \frac{1}{2g^3}\sum_{\mu , \nu \in \mathcal{L}_1, \mu \neq \nu} S_\mu S_\nu \notag\\
  &+ \tfrac{15/16N_S+45/16\sqrt{N_S}+10-(N_S/2+3\sqrt{N_S}/2+4)n_1}{g^3} P_L \,.
\end{align}
(In this expression, $n_1$ is the number of $Z$-errors at sites in $\mathcal{L}_1$ relative to the cluster state.  It appears is this expression because $\theta^{(4)}$ depends on the second-order energies.)
This operator is diagonal in the familiar basis
$\{|C\rangle,|C\{\alpha\}\rangle,|C\{\alpha,\beta\}\rangle,\ldots
\}$.  The corresponding corrections to the lowest energy ground-state energy are
\begin{align}
    \lambda^2 E^{(2)}_{|C\rangle}
    &= -\frac{(N_S+3\sqrt{N_S} +8)}{g}\lambda^2 \,, \notag \\
    \lambda^3 E^{(3)}_{|C\rangle}
    &= -\frac{3\sqrt{N_S}}{2g^2}\lambda^3 \,, \notag \\
    \lambda^4 E^{(4)}_{|C\rangle}
    &= \frac{-6N_S-3\sqrt{N_S}+128}{16}\frac{\lambda^4}{g^3} \,.
\end{align}
Thus, the non-degenerate ground state of the system is the cluster state
$|C\rangle$, to zeroth order, with an energy to fourth-order given by
\begin{align}
    E_0 = &-4gN_S-12g\sqrt{N_S}-8g -\frac{(N_S+3\sqrt{N_S} +8)}{g}\lambda^2 \notag\\
&\quad-\frac{3\sqrt{N_S}}{2g^2}\lambda^3
    -\frac{6N_S+3\sqrt{N_S}-128}{16g^3}\lambda^4 \,.
\end{align}
A state $|C\{\alpha,\beta,\ldots\}\rangle$ obtained from the cluster
state by $n_1$ $Z$-errors at sites in $\mathcal{L}_1$, $n_2$ $Z$-errors at
sites in $\mathcal{L}_2$ and $n_3$ $Z$-errors at sites in $\mathcal{L}_3$ will have
energy
\begin{equation}
    E_{(n_1,n_2,n_3)}
    = E_{|C\rangle} + \Delta(n_1) + n_2 \Delta_2 + n_3 \Delta_3 \,,
\end{equation}
where
\begin{align}
\Delta_3 &:= \frac{5\lambda^4}{8g^3} \qquad  \Delta_2 :=\frac{3\lambda^3}{4g^2} \notag\\
\Delta(n_1) &:=(\frac{2\lambda^2}{g}-\frac{6\lambda^4}{g^3})n_1+\frac{2\lambda^4}{g^3} {n_1}^2 +k_{n_1} \\
 k_0 &:=k_4:=0\,,\quad k_1:= k_3:= \frac{6\lambda^4}{g^3}\,, \quad k_2:=\frac{8\lambda^4}{g^3} \,. \notag
\end{align}
For the range $\lambda/g<1$, the energy gap will be $\Delta_3 =
(5/8)\lambda^4/g^3$, the same energy gap obtained using periodic boundary conditions.

In summary, the perturbative procedure is still successful on the line and square lattice with fixed boundary conditions, producing an approximate cluster state as the non-degenerate ground state.  Furthermore, the energy gap to the first-excited space is unchanged from the case with periodic boundary conditions.  Extending these results to other lattices is straightforward.

\section{The Cluster State on a General Graph}
\label{sec:GeneralGraph}

We have shown that the perturbative procedure presented here is successful in producing a non-degenerate ground state that approximates the cluster state on all of the lattice types examined so far. In fact, the cluster state on any lattice type with any boundary conditions, or more generally on any graph, can be approximated using this method.  We now outline a proof of this result.

For any graph, we place at each site a number of physical qubits equal to the coordination number (the number of bonds connecting that site to others) and take the Hamiltonian as in Eq.~(\ref{hamilt}).  We note that the form of the site Hamiltonian $H_0$ needs only yield a two-dimensional degenerate ground state spanned by $|00\ldots 0\rangle$ and $|11\ldots 1\rangle$ of all qubits at each site; aside from this requirement, its precise form is quite flexible.

We first show that the operators produced at each order in perturbation theory will always possess the cluster state as an eigenstate, and more generally are diagonalized by the set of cluster states with $Z$ errors.  Note that the operators that arise at each order of the perturbation theory are linear combinations of terms of the form
\begin{equation}
  P_L V \bigl[{\textstyle \prod_{k=1}^l} (\Omega^{\alpha_k} \bar{P}_L V)\bigr]P_L\,,
\end{equation}
for some integers $l$ and $\alpha_k$, where $\Omega:=(E_0^{(0)}-gH_0)^{-1}$.  Now, $V$ is a sum of operators $\sigma^z_{(\mu,i)}\otimes \sigma^x_{(\nu,j)}$ where $\mu \neq \nu$, and therefore $P_L V (\prod_{k=1}^l (\Omega^{\alpha_k} \bar{P}_L V)) P_L$ will be a sum of operators which map states in the logical space through the illogical spaces (by applications of $\sigma^z\otimes \sigma^x$ over various bonds) and then return it to the logical space.  From the Pauli operator commutation relations, we note that every term $\sigma^z\otimes \sigma^x$ that is applied to the logical space either commutes or anticommutes with the terms in $H_0$ and so it will always yield an eigenstate of $H_0$.  Thus, successive applications maps the logical space to eigenspaces of definite unperturbed energy, and the term $\Omega=(E_0^{(0)}-gH_0)^{-1}$ will just be a multiplicative constant.

Furthermore, the fact that each application of $\sigma^z\otimes \sigma^x$ keeps the system in some eigenstate of $H_0$ means that all the terms in the sum are of the form $P_L K P_L$, where $K$ is some product of the $\sigma^z\otimes \sigma^x$.  Now, suppose $K$ does not commute with all the terms in $H_0$ (i.e. it anticommutes with at least one of them), then the effect of this term will be that $K$ maps logical states to illogical ones (the resultant state will have a $-1$ eigenvalue for at least one term in $H_0$, whereas the logical space is the $+1$ eigenstate of all the terms in $H_0$).  Thus in this case $P_L K P_L=0$.  The only non-zero operators in the perturbation theory will be of the form $P_L K P_L$ where $K$ commutes with all the terms in $H_0$ (i.e. all the $\sigma^z_{(\mu,i)}\otimes \sigma^z_{(\mu,i+1)}$). But then, if $K$ commutes with each term in $H_0$ it commutes with $P_L$, the projection onto the logical subspace. Thus, we have the eigenvalue relation
\begin{align}
  P_L K P_L |C\rangle &= P_L K P_L \bigl[P_L |C_2\rangle|C_2\rangle\cdots|C_2\rangle\bigr]\notag\\
  &= P_L K |C_2\rangle|C_2\rangle\cdots|C_2\rangle \notag\\
  &= P_L |C_2\rangle|C_2\rangle\cdots|C_2\rangle = |C\rangle
\end{align}
where the last line follows because $\sigma^z\otimes \sigma^x|C_2\rangle=\sigma^x\otimes \sigma^z|C_2\rangle=|C_2\rangle$, and $K$ is a product of $\sigma^z\otimes \sigma^x$ terms.

Hence, we have shown that all the terms that arise at each order in the perturbation theory stabilise the cluster state. This certainly shows that, to zeroth order, the cluster state is one of the eigenstates selected out of the degeneracy by the perturbation. We now show that it is the non-degenerate ground state.

Each term of the form $P_L K P_L$ stabilises the cluster state, and the eigenvalues of $P_L K P_L$ are restricted to $\pm 1$.  Therefore, all that must be checked is that the sign in front of $P_L K P_L$ in the perturbation theory is negative to ensure that the cluster state is selected as the \emph{ground state}.  Suppose $K$ is a product of $m$ $\sigma^z\otimes \sigma^x$ terms. Then the operator $P_L K P_L$ will first appear at the $m$th order as a term in $P_L V [\Omega \bar{P}_L V ]^{m-1} P_L$. The $\Omega$ operators will always contribute a sign $(-1)^{m-1}$ to the energy correction. Furthermore, each $\sigma^z\otimes \sigma^x$ carries with it a negative sign in the definition of $V$, contributing a further $(-1)^m$ to the the energy correction.  Therefore, each $P_L K P_L$ will always appear in the energy correction with a negative sign, thus selecting the cluster state as the ground state. Moreover, because it is clear that the cluster state stabilisers $S_\mu=X_\mu \otimes_{\nu\sim\mu}Z_\nu$, will always arise as one of the $P_L K P_L$ terms in the perturbation theory, the ground state must be non-degenerate (as the state stabilised by these operators is unique). Hence we have shown what we set out to prove: that the cluster state on any lattice type can be approximated using this method. In fact, this perturbative approach can also be further generalised to approximate other states with a PEPS description.

\section{Conclusion}

The existence of gapped quantum many-body systems, with Hamiltonians consisting of only two-body nearest-neighbour interactions, for which the ground state encodes a cluster state allowing universal MBQC is an exciting result for the potential realisation of a quantum computer.  The obvious avenue for future investigation is whether existing natural or artificial materials exist with interactions similar to those described here.

As we have shown, as the perturbation parameter $\lambda/g$ becomes larger, the ground state begins to deviate from the cluster state due to perturbative corrections.  In this work, we have analysed these corrections as a source of error.  It may also be fruitful, however, to analyse the usefulness of the finite $\lambda/g$ ground state for MBQC in terms of the performance of a universal set of quantum gates, as in~\cite{Doh08}.  Although we do not believe the model investigated here exhibits a phase transition at any $\lambda/g$ (this is however an open question), it may nevertheless be possible that the usefulness of the ground state for MBQC undergoes some form of sharp transition~\cite{Bar08}.

\begin{acknowledgments}
  This work was supported by the Australian Research Council.  We thank Sergey Bravyi for identifying that the PEPS Hamiltonian ground state was a simultaneous $+1$ eigenstate of the encoded stabilizer operators $K_\mu$, and for identifying their role in obtaining exact solutions to this model.  We thank Andrew Doherty, Terry Rudolph and Stein Olav Skr{\o}vseth for helpful discussions.
\end{acknowledgments}

\appendix

\section{Perturbation Theory}

We briefly outline the formalism of degenerate perturbation theory and the notation that we use, closely following Ref.~\cite{pt}.

Suppose that the Hamiltonian has the form $H_0+V$, and that the eigenvalue problem has been solved exactly for $H_0$. The corrections brought about by the introduction of the perturbation $V$ can then be approximated by a power series expansion in $V$.  For perturbation theory to converge, the magnitude of the largest eigenvalue of $V$ must be smaller than that of $H_0$.

Suppose the unperturbed spectrum has a degenerate subspace
$\mathcal{H}_L$ with energy $E_L^{(0)}$ and we are interested in finding
out how this energy degeneracy is broken. After the perturbation has
been applied, denote the perturbed eigenstates of this subspace by
$|\psi_i\rangle$ and the perturbed energies by $E_i$, for
$i=1,\ldots,\text{dim}\,\mathcal{H}_L$, i.e.,
\begin{equation}
  \label{A1}
  (H_0+V)|\psi_i\rangle=E_i|\psi_i\rangle \,.
\end{equation}
Denote the projection onto the degenerate subspace $\mathcal{H}_L$
as $P_L$ and define $\overline{P}_L=I-P_L$. Then we can
decompose $|\psi_i\rangle$ as
$|\psi_i\rangle=|\psi_i\rangle_L+|\psi_i\rangle_{\bar{L}}$,
where $|\psi_i\rangle_L:=P_L|\psi_i\rangle$ and
$|\psi_i\rangle_{\bar{L}}:=\overline{P}_L|\psi_i\rangle$.

Applying this decomposition to Eq.~(\ref{A1}), we have:
\begin{equation}
\label{A2}
 (H_0+V)|\psi_i\rangle_L+(H_0+V)|\psi_i\rangle_{\bar{L}}=E_i |\psi_i\rangle_L+E_i|\psi_i\rangle_{\bar{L}}\,.
\end{equation}
Note that $P_L H_0 = H_0 P_L = E_L^{(0)} P_L$ and therefore that
$\overline{P}_L H_0 = H_0 \overline{P}_L$.  Multiplying Eq.~(\ref{A2}) by $P_L$ and
$\overline{P}_L$ respectively, we obtain
\begin{align}
  \label{A3}
  (E_i-E_L^{(0)}-P_L V P_L)|\psi_i\rangle_L - (P_L V \overline{P}_L)|\psi_i\rangle_{\bar{L}}&=0\, \\
  \label{A4}
  (E_i-H_0-\overline{P}_L V \overline{P}_L)|\psi_i\rangle_{\bar{L}} - (\overline{P}_L V P_L)|\psi_i\rangle_L&=0\,.
\end{align}
Eq.~(\ref{A4}) has the formal solution
\begin{equation}
  \label{A5}
  |\psi_i\rangle_{\bar{L}} =(E_i-H_0-\overline{P}_L V
   \overline{P}_L)^{-1}(\overline{P}_L V P_L)|\psi_i\rangle_L\,,
\end{equation}
which can be substituted back into Eq.~(\ref{A3}) to obtain
\begin{equation}
\label{A6} \theta |\psi_i\rangle_L=(E_i-E_L^{(0)})|\psi_i\rangle_L\,,
\end{equation}
where
\begin{multline}
  \theta:=P_L V P_L \\
  +(P_L V
  \overline{P}_L)(E_i-H_0-\overline{P}_L V
  \overline{P}_L)^{-1}(\overline{P}_L V P_L)\,.
\end{multline}
This equation allows us to determine perturbed energy at any order of the perturbation
theory.  So far no approximations have been made.  To implement the
perturbation theory, it is necessary to expand $\theta$ as a power
series in $V$.  We use
\begin{align}
  \label{A8}
  (E_i-&H_0-\overline{P}_L V \overline{P}_L)^{-1}\notag\\
&=\bigl(I-(E_i-H_0)^{-1}\overline{P}_L V \overline{P}_L\bigr)^{-1}(E_i-H_0)^{-1}\notag\\
&=\Bigl(\sum_{m=0}^{\infty}\bigl[(E_i-H_0)^{-1}\overline{P}_L V
\overline{P}_L\bigr]^{m}\Bigr)(E_i-H_0)^{-1}\,.
\end{align}
The energies $E_i$ in these expressions must also be expanded as power series,
$E_i=E_L^{(0)}+\sum_{k=1}^{\infty}E_i^{(k)}$. Then we have
\begin{align}
\label{A9}
(E_i-&H_0)^{-1}=(E_L^{(0)}+\sum_{k=1}^{\infty}E_i^{(k)}-H_0)^{-1}\notag\\
&=\Bigl[I+(E_L^{(0)}-H_0)^{-1}\sum_{k=1}^{\infty}E_i^{(k)}\Bigr]^{-1}(E_L^{(0)}-H_0)^{-1}\notag\\
&=\Lambda\Omega
\end{align}
where we have defined the operators
\begin{align}
  \Omega &:=(E_L^{(0)}-H_0)^{-1}\,, \\
  \Lambda &:=\bigl(I+\Omega\sum_{k=1}^{\infty}E_i^{(k)}\bigr)^{-1}\,.
\end{align}
The operator $\theta$ can then be expressed as
\begin{equation}
  \label{A10}
  \theta=P_L V P_L + P_L V
  \sum_{m=0}^{\infty}[\Lambda\Omega
  \overline{P}_L V ]^{m}
  \Lambda\Omega \overline{P}_L V P_L\,,
\end{equation}
where we have used the fact that
$\overline{P}_L$ commutes with $\Omega$ and $\Lambda$. Additionally, $\Lambda$ must be further
expanded out as a power series
\begin{equation}
  \label{A10l}
  \Lambda=(I+\Omega\sum_{k=1}^{\infty}E_i^{(k)})^{-1}=\sum_{j=0}^{\infty}(-\Omega\sum_{k=1}^{\infty}E_i^{(k)})^{j}\,.
\end{equation}

We can now identify the terms in Eq.~(\ref{A10}) of each order. Specifically, denote the terms in $\theta$ of $k$th order by $\theta^{(k)}$, so that $\theta=\sum_{k=1}^{\infty}\theta^{(k)}$. Then, when approximated to $n$th order in $V$, Eq.~(\ref{A6}) becomes
\begin{equation}
  \label{A7}
  \sum_{k=1}^{n}\theta^{(k)} |\psi_i\rangle_L=\sum_{k=1}^{n}E_i^{(k)}|\psi_i\rangle_L\,,
\end{equation}
which is an eigenvalue equation over the subspace $L$. (We note that
$|\psi_i\rangle_L \neq 0$ because $|\psi_i\rangle \in \mathcal{H}_L$
at zeroth order and the perturbation is assumed small).  The
energy corrections to $n$th order $\sum_{k=1}^{n}E_i^{(k)}$ are
the eigenvalues of the operator $\sum_{k=1}^{n}\theta^{(k)}$.
Furthermore, the eigenstates that are selected by the perturbation
to break the degeneracy (to zeroth order) are just the eigenvectors
of $\sum_{k=1}^{n}\theta^{(k)}$ corresponding to each eigenvalue.
Note that $\theta^{(k)}$ depends on lower order energy corrections
and so each lower order correction must be calculated before
proceeding to the higher order corrections. At each stage we must insist that the $n$th order
energies differ from the $(n{-}1)$th order energies only by an
amount $n$th order in $V$, which removes any non-physical solutions.

To determine the explicit form of the $\theta^{(k)}$, we simply substitute Eq.~(\ref{A10l}) into Eq.~(\ref{A10}) and identify the terms of the required order.
Clearly $\theta^{(1)}=P_L V P_L$, and so to first order in $V$, Eq.~(\ref{A7}) becomes
\begin{equation}
  \label{A11}
  \theta^{(1)} |\psi_i\rangle_L=P_L V P_L |\psi_i\rangle_L=E_i^{(1)}|\psi_i\rangle_L\,.
\end{equation}
Thus, the first order energy corrections to states in $\mathcal{H}_L$
are just the eigenvalues of the matrix $P_L V P_L$. In the
non-degenerate case we see that Eq.~(\ref{A11}) reduces to the well-known expression $E_i^{(1)}=\langle L| V|
L\rangle$. If this eigenvalue spectrum is still degenerate, then the
degeneracy is not completely broken at first order. It is then
necessary to go to second order
\begin{align}
  \label{A12}
  (E_i^{(1)}+E_i^{(2)})|\psi_i\rangle_L&=(\theta^{(1)}+\theta^{(2)}) |\psi_i\rangle_L\notag\\
  &=(P_L V P_L+P_L V \Omega \overline{P}_L V P_L) |\psi_i\rangle_L\,.
\end{align}
Again we can examine whether the degeneracy has been broken at this
stage by examining the eigenvalues and eigenvectors of the above
operator $P_L V P_L + P_L V \Omega \overline{P}_L V P_L$. If not, we
must continue to proceed to higher orders. The formulae for higher
orders become increasingly complex, but they simplify if we assume
that $E_i^{(1)}=0$, which will be true for all the cases of interest
that we investigate in this paper.  For example we find (with $E_i^{(1)}=0$)
\begin{align}\label{opert}
\theta^{(1)}&= P_L V P_L\notag\\
\theta^{(2)}&= P_L V \Omega \overline{P}_L V P_L\notag\\
\theta^{(3)}&= P_L V [\Omega \overline{P}_L V]^2 P_L \notag\\
\theta^{(4)}&= P_L V [\Omega \overline{P}_L V ]^3 P_L- E_i^{(2)} P_L V\Omega^2 \overline{P}_L V P_L\,,
\end{align}
and so forth.

There are two additional points that must be noted.  First, from Eq.~(\ref{A10}), it can be concluded that $\sum_{k=1}^{N}E_i^{(k)}$ are
the eigenvalues of the operator $\sum_{k=1}^{N}\theta^{(k)}$ for each order $N$. However, the energies $E_i^{(k)}$ are not generally the eigenvalues of the operators $\theta^{(k)}$; this will only be true in general when all operators $\theta^{(k)}$ can be
simultaneously diagonalised.  Fortunately, for the systems investigated in this paper, it can be shown that the $\theta^{(k)}$ commute with each other and therefore the energies $E_i^{(k)}$ are indeed the eigenvalues of the operators $\theta^{(k)}$.
Second, there are some general properties we can note about the form of the $\theta^{(k)}$ operators. From the form of Eq.~(\ref{A10}) and the series expansion in Eq.~(\ref{A10l}), it is clear that all terms in the expansion of $\theta$ are proportional to
operators of the form $P_L V (\prod_{k=1}^l (\Omega^{\alpha_k} \overline{P}_L V)) P_L$ for some $l, \alpha_k \in \mathbb{N}$. This result is used in Sec.~\ref{sec:GeneralGraph}.

The above analysis allows the zeroth order energy eigenstates to be
determined.  We now direct our attention to the first order corrections to these states.  Suppose we have
determined that $|\psi_i\rangle=|\psi_i\rangle_L=|i\rangle$ to
zeroth order for some $|i\rangle \in \mathcal{H}_L$ using the above
method. From Eq.~(\ref{A5}) and using Eqs.~(\ref{A8})
and (\ref{A9}) we have, to first order in $V$,
\begin{equation}
  \label{A15}
  |\psi_i\rangle_{\bar{L}} = \Omega (\overline{P}_L V P_L)|\psi_i\rangle_L =
  \sum_{|j\rangle \in \mathcal{H}_{\bar{L}}} \frac{ \langle j | V |i\rangle}{ E_L^{(0)}-E_j^{(0)}}|j\rangle\,,
\end{equation}
where $|j\rangle \in \mathcal{H}_{\bar{L}}$ are the eigenstates of
$H_0$ with energy $E_j^{(0)}$. Determining $|\psi_i\rangle_L$ to first order is somewhat
more complicated: even though $|\psi_i\rangle_L=|i\rangle$ to zeroth
order, it is possible for first order corrections to come from other
states $|l\rangle \in \mathcal{H}_L$. However, if the energy
degeneracy between $|i\rangle$ and $|l\rangle$ is only broken at
order $m_l$, then one must go to the order $(m_l+1)$
equations just to determine the \emph{first} order eigenstate
corrections to $|\psi_i\rangle_L$. In this case,
$\sum_{k=1}^{m_l-1}E_i^{(k)}=\sum_{k=1}^{m_l-1}E_l^{(k)}$ but
$E_i^{(m_l)} \neq E_l^{(m_l)}$. Then, at order $(m_l+1)$,
Eq.~(\ref{A7}) reads
\begin{equation}
  \label{A16}
  (\sum_{k=1}^{m_l}\theta^{(k)}+\theta^{(m_l+1)})
  |\psi_i\rangle_L=\sum_{k=1}^{m_l+1}E_i^{(k)}|\psi_i\rangle_L\,.
\end{equation}
Taking the inner product with $|l\rangle$ and rearranging gives
\begin{align}
\langle l |\psi_i\rangle_L&=\frac{\langle l |\theta^{(m_l+1)} |\psi_i\rangle_L -E_i^{(m_l+1)}\langle l |\psi_i\rangle_L}{E_i^{(m_l)}-E_l^{(m_l)}} \notag\\
&=\frac{\langle l |\theta^{(m_l+1)}
|i\rangle}{E_i^{(m_l)}-E_l^{(m_l)}}\,,
\end{align}
to first order in $V$.  Therefore, to first order in $V$, we have
\begin{multline}
\label{A19} |\psi_i\rangle =  |i\rangle + \sum_{|j\rangle \in
\mathcal{H}_{\bar{L}}} \frac{ \langle j | V |i\rangle}{
E_L^{(0)}-E_j^{(0)}}|j\rangle \\
+ \sum_{|l\rangle \in \mathcal{H}_L, l
\neq i} \frac{\langle l |\theta^{(m_l+1)}
|i\rangle}{E_i^{(m_l)}-E_l^{(m_l)}}|l\rangle\,.
\end{multline}

\end{document}